\documentclass[aps, prl, reprint, superscriptaddress]{revtex4-1}
\usepackage[english]{babel}
\usepackage{amsmath,amsthm}
\usepackage{amsfonts}
\usepackage{xspace}
\usepackage{bm}
\usepackage{booktabs}
\usepackage{graphicx}
\usepackage{dcolumn}
\usepackage[mathlines]{lineno}
\usepackage[colorlinks]{hyperref}  
\usepackage{gensymb}
\usepackage{epstopdf}
\usepackage{newfloat}
\usepackage[usenames, dvipsnames]{color}
\DeclareFloatingEnvironment[name={Figure S}]{suppfigure}
\DeclareFloatingEnvironment[name={S}]{suppEquation}

\emergencystretch=\maxdimen
\newcommand{\tNiFe}{\ensuremath{t_\mathrm{NiFe}}\xspace}
\newcommand{\tIrMn}{\ensuremath{t_\mathrm{IrMn}}\xspace}
\newcommand{\tCoFe}{\ensuremath{t_\mathrm{CoFe}}\xspace}

\newcommand{\Hex}{\ensuremath{H_\mathrm{ex}}\xspace}
\newcommand{\Hsat}{\ensuremath{H_\mathrm{sat}}\xspace}

\newcommand{\Ms}{\ensuremath{M_\mathrm{s}}\xspace}

\newcommand{\DelH}{\ensuremath{\Delta H}\xspace}
\newcommand{\lsd}{\ensuremath{\lambda_\mathrm{sd}}\xspace}

\newcommand{\geff}{\ensuremath{g^{\uparrow\downarrow}_\mathrm{eff}}\xspace}

\begin{document}

\title{Spin decoherence independent of antiferromagnetic order in IrMn}

\author{Behrouz Khodadadi}%
\affiliation{ 
Department of Physics, Virginia Tech, Blacksburg, VA 24061, USA
}
\author{Youngmin Lim}%
\affiliation{ 
Department of Physics, Virginia Tech, Blacksburg, VA 24061, USA
}
\author{David A. Smith}%
\affiliation{ 
Department of Physics, Virginia Tech, Blacksburg, VA 24061, USA
}
\author{Ryan W. Greening}%
\affiliation{ 
Department of Physics, Virginia Tech, Blacksburg, VA 24061, USA
}
\author{Yuankai Zheng}%
\affiliation{ 
Western Digital, Fremont, CA 94539, USA
}
\author{Zhitao Diao}%
\affiliation{ 
Western Digital, Fremont, CA 94539, USA
}

\author{Christian Kaiser}%
\affiliation{ 
Western Digital, Fremont, CA 94539, USA
}
\author{Satoru Emori}%
\affiliation{ 
Department of Physics, Virginia Tech, Blacksburg, VA 24061, USA
}
\email
{semori@vt.edu}

\date{December 3, 2018}

\begin{abstract}

We investigate the impact of pinned antiferromagnetic order on the decoherence of spin current in polycrystalline IrMn. 
In NiFe/Cu/IrMn/CoFe multilayers, we coherently pump an electronic spin current from NiFe into IrMn, whose antiferromagnetic order is globally pinned by static exchange-bias coupling with CoFe. We observe no anisotropic spin decoherence with respect to the orientation of the pinned antiferromagnetic order.
We also observe no difference in spin decoherence for samples with and without pinned antiferromagnetic order. 
Moreover, although there is a pronounced resonance linewidth increase in NiFe that coincides with the switching of IrMn/CoFe, we show that this is not indicative of anisotropic spin decoherence in IrMn.
Our results demonstrate that the decoherence of electron-mediated spin current is remarkably insensitive to the magnetization state of the antiferromagnetic IrMn spin sink. 

\end{abstract}
\maketitle

A spin current is said to be coherent when the spin polarization of its carriers, e.g., electrons, is locked in a uniform precessional phase. How a spin current loses its coherence, particularly as it interacts with magnetic order, is a crucial fundamental question in spintronics~\cite{Wolf2001a}. In a ferromagnetic metal (FM), an electronic spin current polarized transverse to the magnetization dephases quickly in the uniform ferromagnetic exchange field~\cite{Ralph2008, Stiles2002}. Experiments of ferromagnetic resonance (FMR) spin pumping~\cite{Tserkovnyak2002, Taniguchi2008, Boone2013}, where a coherently excited spin current propagates from a FM spin source to a FM spin sink~\cite{Li2016c}, show the transverse spin coherence length in FMs to be as short as $\approx$1 nm~\cite{Ghosh2012}. The dephasing of transverse spin polarization $\mathbf{s}$ also gives rise to a spin-transfer torque,  $\propto \mathbf{m}\times\mathbf{s}\times\mathbf{m}$, acting on the magnetization $\mathbf{m}$ of the FM spin sink~\cite{Ralph2008, Stiles2002,Brataas2012,Taniguchi2008a}.

For antiferromagnetic metals (AFMs) with staggered exchange fields, a fundamental understanding of spin transport has yet to be developed by experiment. Although the transverse spin coherence length can in principle be $\gg$1~nm~\cite{Nunez2006, Saidaoui2014, Baltz2018}, an electronic spin current polarized transverse to the antiferromagnetic order parameter (N\'eel vector $\mathbf{l}$) is expected to dephase in the diffusive limit of transport~\cite{Saidaoui2014,Manchon2017}. Such spin dephasing in AFMs generates a spin-transfer torque, $\propto \mathbf{l}\times\mathbf{s}\times\mathbf{l}$~\cite{Gomonay2010,Manchon2017,Baltz2018}, which may be crucial for emerging antiferromagnetic spintronic technologies~\cite{Jungwirth2016, MacDonald2011,Wadley2016,Bodnar2018, Chen2018}. Furthermore, spin dephasing in an AFM with a \emph{uniform} N\'eel vector may yield anisotropic decoherence, where spin absorption by the AFM is enhanced when $\textbf{l} \perp \textbf{s}$~\cite{Moriyama2017}.

By contrast, \emph{polycrystalline} thin films of AFMs by themselves do not exhibit anisotropic spin decoherence on a macroscopic scale, since the grains contain a distribution of N\'eel vector orientations that averages out the anisotropy ~\cite{Zhang2014}.  
While polycrystalline AFMs have found commercial applications (i.e., pinning ferromagnetic layers in spin valves)~\cite{Gider1998}  and been used as spin sinks ~\cite{Ghosh2012, Merodio2014,Zhang2014,Qu2015,Mendes2014,Klaui2018,Frangou2016}, 
their nonuniform, unpinned antiferromagnetic order poses a challenge for gaining fundamental insight into spin decoherence.

To align the global antiferromagnetic order, a polycrystalline AFM can be exchange-bias-coupled to a ferromagnetic metal (FM) ~\cite{Stamps2000,Wei2007,Urazhdin2007}; the N\'eel vector \textbf{l} can be pinned along the direction of the bias field during film deposition or field cooling.  
In such exchange-biased polycrystalline FM/AFM bilayers, a recent spin pumping experiment has reported anisotropic relaxation of pure spin current in the AFM layer governed by dephasing, i.e., spin transfer acting on \textbf{l}~\cite{Moriyama2017}. This claim is based on the observation of higher magnetic damping when the FM is magnetized away from the exchange bias direction, generating misalignment between the time-averaged \textbf{s} and \textbf{l}~\cite{Moriyama2017}. 
However, the direct interface between the FM spin source and the AFM spin sink introduces interlayer magnon coupling, which may yield similar anisotropic damping due to two-magnon scattering 
within the FM~\cite{McMichael1998,Mewes2010,BeikMohammadi2017}. For FM/AFM bilayers, it is therefore difficult to discern whether pumped spin is transferred to the AFM or decoheres within the FM. 
Moreover, the spin current in metallic FM-spin-source/AFM-spin-sink bilayers can be carried by both electrons and magnons    ~\cite{Baltz2018,Tserkovnyak2002,Rezende2016}, potentially further complicating the interpretation of the spin pumping experiment. 

\par
Here, we investigate how pinned antiferromagnetic order impacts the decoherence of spin current pumped into polycrystalline AFM IrMn. 
We leverage spin-valve-like multilayers of NiFe/Cu/IrMn/CoFe, containing separate FMs for pumping the spin current (NiFe) and for exchange biasing (CoFe). This way, the N\'eel vector in the IrMn spin sink is aligned by exchange bias coupling with CoFe, while the pure spin current from NiFe to IrMn propagating through the diamagnetic Cu spacer is carried entirely by electrons, i.e., without direct magnon coupling. 
From spin pumping measurements of the multilayers, we find that spin decoherence within IrMn is not impacted in any detectable way by (1) the relative alignment between the pumped spin polarization and antiferromagnetic order or (2) the presence of pinned antiferromagnetic order. Our findings indicate that the decoherence of electronic spin current is remarkably insensitive to the magnetization state of AFM IrMn. 

 \begin{figure}[t!]
\includegraphics[width=0.99\columnwidth]{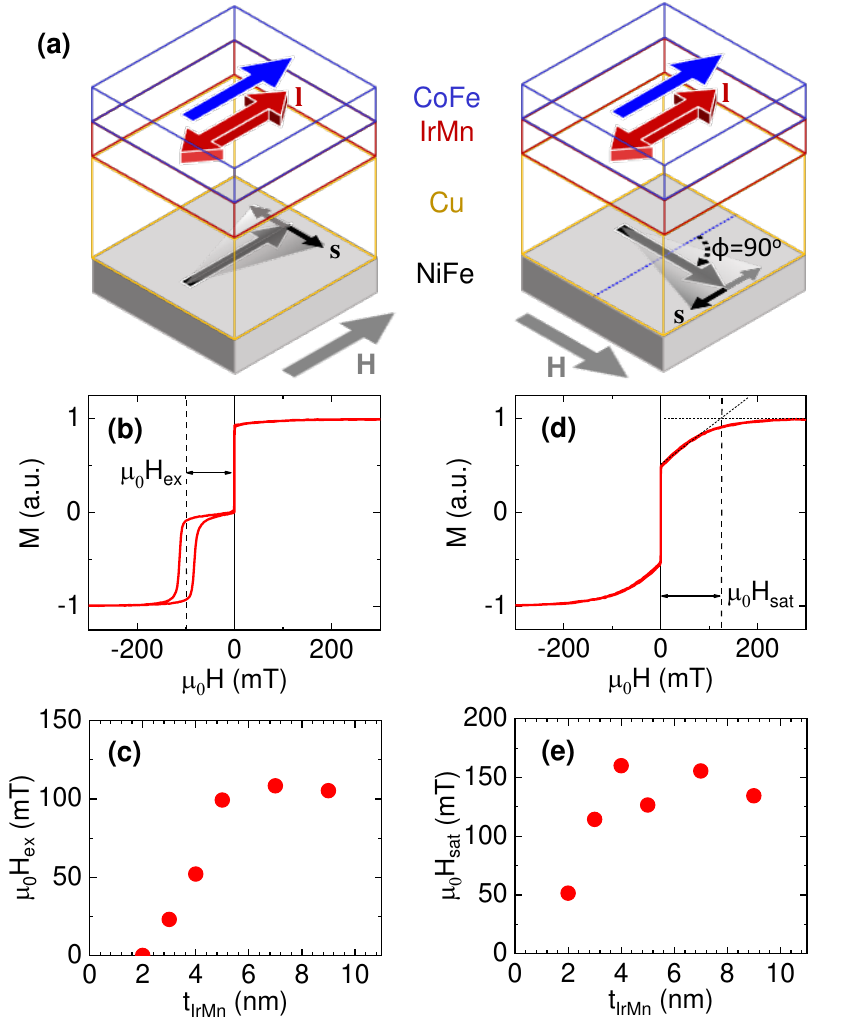}
\caption{\label{fig:static} (a) Illustrations of the multilayer with the soft NiFe magnetization oriented collinear ($\phi = 0^\circ$) and orthogonal ($\phi = 90^\circ$) to the N\'eel vector \textbf{l} in the exchange-biased IrMn layer. The dominant transverse spin polarization \textbf{s}, pumped by FMR in NiFe, is also shown. (b) Magnetization curve for NiFe/Cu/IrMn(5)/CoFe with field applied collinear to the exchange bias direction. $H>0$ corresponds to $\phi = 0^\circ$ whereas $H<0$ to $\phi = 180^\circ$. (c) Exchange bias field \Hex as a function of IrMn thickness \tIrMn. (d,e) Representative magnetization curve (d) and saturation field \Hsat of the CoFe layer (e) for field applied orthogonal ($\phi = 90^\circ$) to the exchange bias direction.}
\end{figure}

\par
The multilayers were fabricated using an Anelva C7100 magnetron sputtering tool with the stack structure of Si/SiO$_2$(substrate)/Ta(3)/Ru(3)/Ni$_{80}$Fe$_{20}$(8)/Cu(4)/\linebreak Ir$_{20}$Mn$_{80}$(\tIrMn)/Co$_{75}$Fe$_{25}$(\tCoFe)/Ru(3)/Ta(3)/Ru(2), where each number in parentheses indicates the layer thickness in nm. 
All samples were annealed at 300$^\circ$C under an in-plane field of 5 T and then cooled to room temperature under the same field. 
Stacks without CoFe (\tCoFe = 0) were made as control samples with a random distribution of N\'eel vectors in IrMn, whereas those with \tCoFe = 4 nm were made to pin the global N\'eel vector orientation by exchange bias. 
All subsequent measurements were performed at room temperature. 

In the multilayer structure as illustrated in Fig.~\ref{fig:static}(a)), the magnetization of soft NiFe (spin source) can be rotated independently to angle $\phi$ from the orientation of the fixed N\'eel vector in AFM IrMn (spin sink), which is pinned by the exchange bias coupling with the hard CoFe layer.
Vibrating sample magnetometry was performed with a Microsense EZ9 VSM to characterize the static magnetic properties of the multilayers. The saturation magnetization \Ms is $\approx$800 kA/m for NiFe and $\approx$1600 kA/m for CoFe.
Magnetometry with field applied collinear to the annealing field direction reveals that the effective exchange bias field reaches $\approx$100 mT, as quantified by the shift of the CoFe hysteresis loop (Fig.~\ref{fig:static}(b)). As summarized in Fig.~\ref{fig:static}(c), finite exchange bias only becomes evident for NiFe/Cu/IrMn/CoFe samples with $\tIrMn \geq 3$ nm, consistent with previous reports~\cite{Ali2003,Chen2014}. 
With field applied orthogonal to the exchange bias direction, while NiFe is essentially saturated at $\lesssim$1 mT, CoFe does not saturate up to $\gtrsim$100 mT (Fig.~\ref{fig:static}(d,e)). 
The static magnetometry thus indicates that at in-plane applied fields $\ll$100 mT, the precessional axis of the pumped spin current, parallel to the static equilibrium magnetization in NiFe~\cite{Tserkovnyak2002}, can be rotated while the N\'eel vector \textbf{l} in IrMn remains mostly aligned along the exchange bias direction.   

\begin{figure}[t!]
\includegraphics[width=0.99\columnwidth]{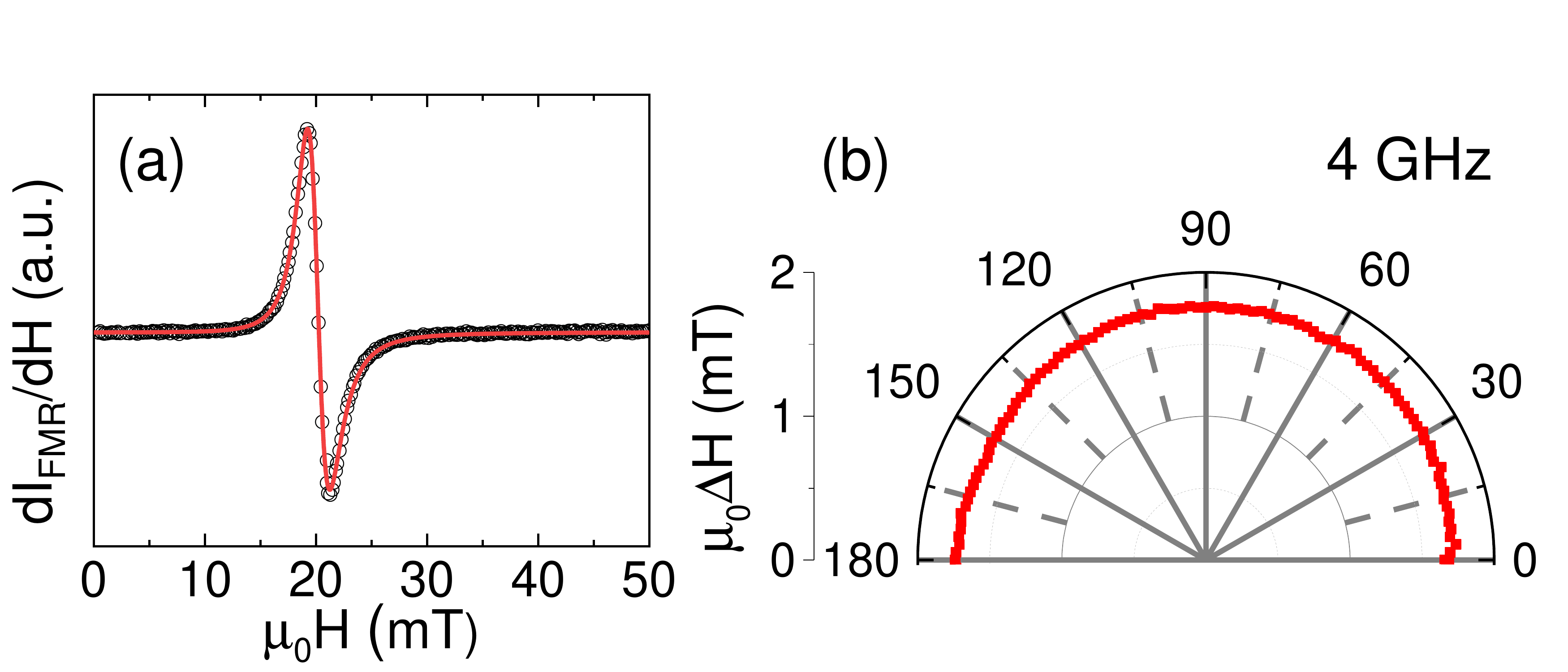}
\caption{\label{fig:angle} (a) Exemplary FMR spectrum at $f = 4$ GHz of NiFe/Cu/IrMn(5)/CoFe. Only the FMR peak from NiFe is observed in the measured range. (b) Dependence of FMR linewidth on the angle $\phi$ between the NiFe magnetization direction and the IrMn/CoFe exchange bias direction.} 
\end{figure}

\par
To test whether the decoherence of spin current is influenced by its relative orientation with the N\'eel vector, we performed angular dependent FMR measurements in the field range well below the orthogonal saturation field of the CoFe (Fig.~\ref{fig:static}(d)).
The orientation between the NiFe magnetization (spin polarization) and the N\'eel vector was set by rotating the film sample attached to a stepper motor with respect to the external quasistatic magnetic field and rf coplanar waveguide. 
Figure~\ref{fig:angle}(a) shows a representative FMR spectrum at 4 GHz with resonance centered at $\mu_0H \approx 20$ mT. 
Around this field, when the NiFe layer magnetized orthogonal ($\phi = 90^\circ$) to the exchange bias direction, the misalignment between the spin precession axis and the IrMn N\'eel vector \textbf{l} is estimated to be $\cos^{-1}(H/\Hsat) \approx 80^\circ$.
We note that in this low frequency regime ($\lesssim$10 GHz), CoFe with high \Ms does not exhibit FMR, such that only the FMR response from NiFe is detected (Fig.~\ref{fig:angle}(a))~\cite{Note1}.

\par
We monitor the decoherence of spin current pumped out of NiFe by measuring the FMR linewidth \DelH ~\cite{Tserkovnyak2002}, defined as the half-width-at-half-maximum from the Lorentzian derivative fit (e.g., Fig.~\ref{fig:angle}(a)).  
Here, the dominant transverse spin polarization \textbf{s} is in the film plane, as illustrated in Fig.~\ref{fig:static}(a), due to the elliptical precessional orbit from thin-film shape anisotropy. One might thus hypothesize anisotropic spin dephasing, such that \DelH is enhanced at $\phi = 0^\circ$, i.e., when \textbf{s} is mostly orthogonal to \textbf{l}~\cite{Moriyama2017}. Alternatively, one might equate \textbf{s} to the time-averaged pumped spin polarization~\cite{Tserkovnyak2002, Moriyama2017}, parallel to the NiFe magnetization, in which case \DelH would be enhanced at $\phi = 90^\circ$.
However, as shown in Fig.~\ref{fig:angle}(b), angular dependent measurements reveal no variation in \DelH.
This result suggests that the decoherence of the pumped spin current is invariant with the pinned antiferromagnetic order in the IrMn spin sink.  

\par
\begin{figure}[tb]
\includegraphics[width=1\columnwidth]{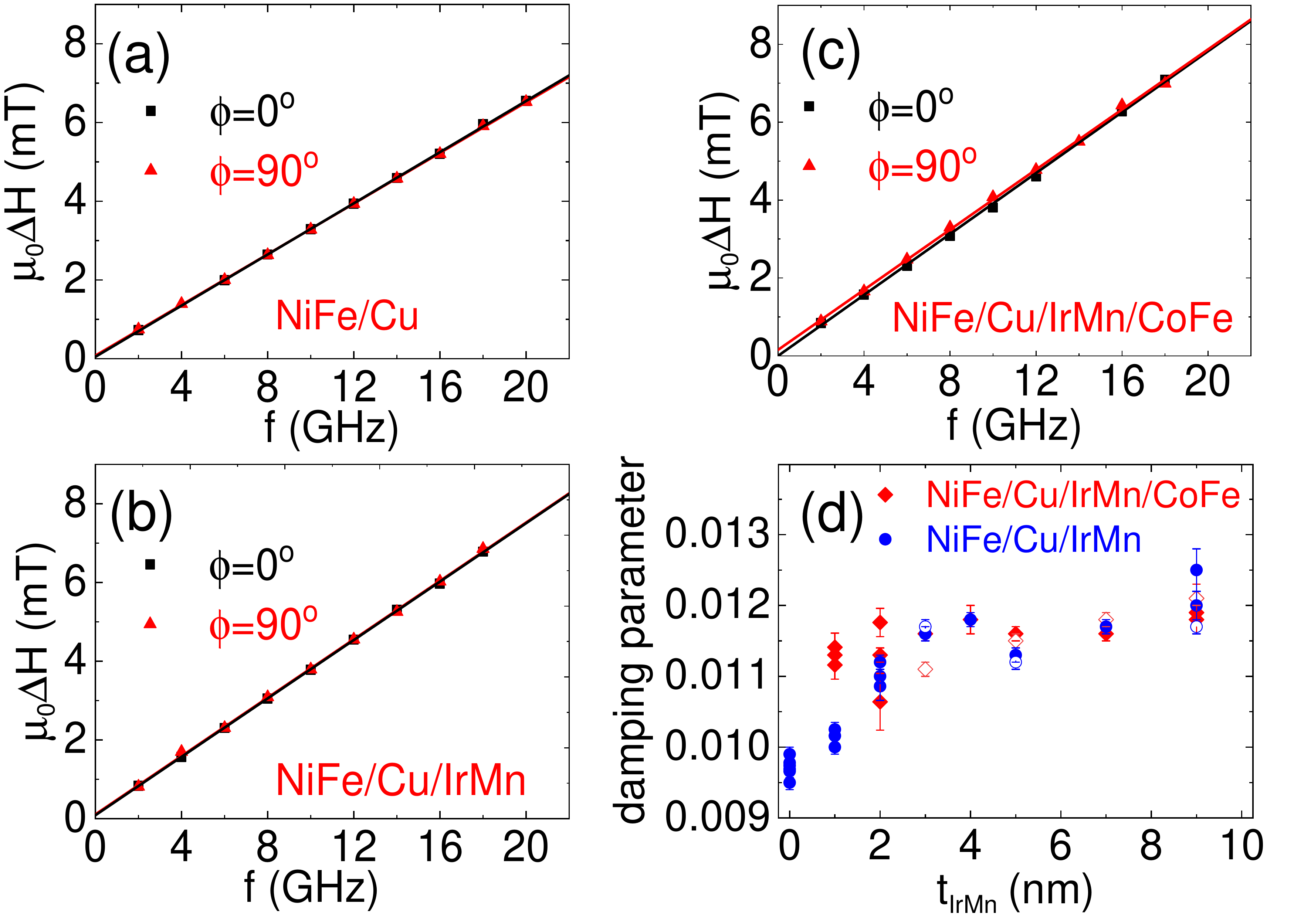}
\caption{\label{fig:Gilbert} Frequency dependence of FMR linewidth with field applied along ($\phi = 0^\circ$) or orthogonal ($\phi = 90^\circ$) to the exchange bias field for (a) NiFe/Cu (b) NiFe/Cu/IrMn(5) (c) NiFe/Cu/IrMn(5)/CoFe. (d) The dependence of the Gilbert damping parameter $\alpha$ on IrMn thickness \tIrMn when field is applied along the exchange bias (solid symbols) or orthogonal (hollow symbols) for NiFe/Cu/IrMn and NiFe/Cu/IrMn/CoFe samples.}
\end{figure}

As shown in Fig.~\ref{fig:Gilbert}(a-c), we quantify the Gilbert damping parameter $\alpha$ from the linear dependence of \DelH on frequency $f$, i.e., $\DelH = \DelH_0 + 2\pi\alpha f/(\mu_0\gamma)$,
where $\DelH_0$ is the zero-frequency linewidth and $\gamma/(2\pi) \approx 29.5$ GHz/T. 
Enhanced damping of up to $\Delta\alpha \approx 0.002$ is observed for samples with IrMn(/CoFe) spin sinks, compared to the sample with \tIrMn = 0 (Fig.~\ref{fig:Gilbert}(a)). 
This confirms the role of the IrMn layer as a spin sink with an effective spin-mixing conductance at the Cu/IrMn interface of $\geff = 4\pi\Ms\tNiFe\Delta\alpha/(\gamma\hbar) \approx 8$ nm$^{-2}$, quantitatively similar to Refs.~\cite{Ghosh2012,Merodio2014}.
Remarkably, we do not observe any significant difference in the frequency dependence of linewidth when NiFe is magnetized parallel ($\phi = 0^\circ$) or orthogonal ($\phi = 90^\circ$) to the annealing field (exchange bias) direction. While AFM IrMn absorbs the pumped spin current from NiFe, spin decoherence is evidently not affected by the relative orientation between the spin current polarization and the global N\'eel vector. 

Another remarkable observation is that spin decoherence is not affected by whether or not IrMn is exchange-biased.  
Figure~\ref{fig:Gilbert}(d) plots the Gilbert damping parameter against \tIrMn for the series of samples with and without CoFe that pins the N\'eel vector of IrMn. 
The damping parameter is essentially identical at $\alpha \approx 0.011-0.012$ for both series with $\tIrMn \gtrsim 2$ nm regardless of whether exchange-bias pinning is present in IrMn. 
Our findings in Figs.~\ref{fig:angle} and \ref{fig:Gilbert} point to the absence of the global anisotropic dephasing. 
The decoherence of spin current in AFM IrMn must be dominated by a mechanism independent of the N\'eel vector, e.g., due to spin-orbit interactions as we discuss later. 

The significantly lower damping in the \tIrMn = 1 nm sample without CoFe suggests that the spin current is not fully absorbed in the nominally 1-nm-thick IrMn layer, which is consistent with prior reports~\cite{Ghosh2012,Merodio2014}. By contrast, the damping exhibits the saturated value for the \tIrMn = 1 nm sample with CoFe, because the 4-nm-thick ferromagnetic CoFe layer fully absorbs the spin current~\cite{Ghosh2012}. 

\begin{figure}[tb]
\includegraphics[width=1.0\columnwidth]{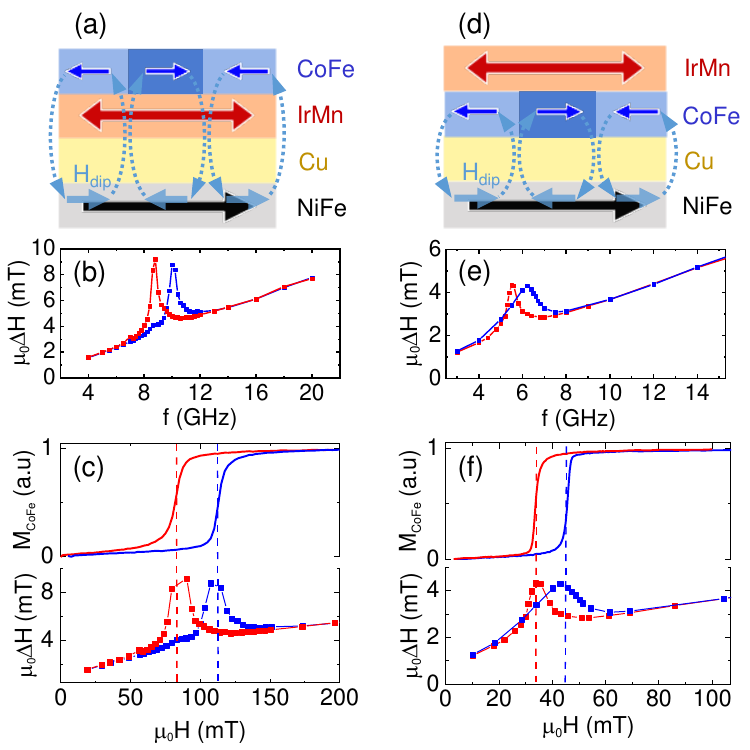}
\caption{\label{fig:LWpeaks} (a,d) Cross-sectional illustrations of the (a) NiFe/Cu/IrMn/CoFe and  (d) NiFe/Cu/CoFe/IrMn stack structures. The local dipole fields $H_\mathrm{dip}$ arise only when the exchange-biased CoFe layer undergoes switching.
(b,e) Frequency dependence of FMR linewidth obtained with the high-to-low (red) and low-to-high (blue) field sweeps, with the field direction ($\phi = 180^\circ$) opposite to the exchange bias direction, for (b) NiFe/Cu/IrMn(5)/CoFe and (e) NiFe/Cu/CoFe/IrMn(5). 
(c,f) Comparison of the switching of CoFe magnetization $M_\mathrm{CoFe}$ and the field dependence of FMR linewidth for (c) NiFe/Cu/IrMn(5)/CoFe and (f) NiFe/Cu/CoFe/IrMn. The dashed vertical lines are guides to the eye that show CoFe switching and FMR linewidth peaks coincide.}  
\end{figure}

From Figs.~\ref{fig:angle} and \ref{fig:Gilbert}, we find that the decoherence of spin current in IrMn, probed through the damping of NiFe, does not change when rotating the pumped spin polarization with respect to the exchange bias direction. 
However, as shown in Fig.~\ref{fig:LWpeaks}(a,b), we find a pronounced enhancement in FMR linewidth within a narrow range of frequencies when NiFe is magnetized opposite ($\phi = 180^\circ$) to the exchange bias direction. This anomalous linewidth enhancement is maximized at different frequencies depending on the field sweep protocol during the acquisition of FMR spectra. As shown by the red and blue data points in Fig.~\ref{fig:LWpeaks}(b), the linewidth peak is observed at a higher (lower) frequency when the field is swept from low to high (high to low). 
This hysteretic behavior suggests that the linewidth enhancement is related to the magnetic hysteresis of the exchange-biased CoFe layer. 
Indeed, by comparing the field dependence of linewidth with magnetometry data (Fig.~\ref{fig:LWpeaks})(c), we find that the linewidth enhancement coincides with the switching of CoFe. 

We consider two possible mechanisms for this anomalous linewidth enhancement: (1) the decoherence of spin current in IrMn via dephasing is enhanced when the N\'eel vector, pinned to the CoFe magnetization, twists between $\phi = 0^\circ$ and $180^\circ$, similar to the mechanism proposed in Ref.~\cite{Moriyama2017}, or (2) the linewidth for NiFe resonance increases due to inhomogeneous dipolar magnetic fields from CoFe in a multidomain state, where FMR modes precessing about different local effective fields overlap to yield a broad linewidth~\cite{Platow1998,McMichael2003,Twisselmann2003}. 
To distinguish between the two mechanisms, we examine FMR and magnetometry measurements performed on a ``conventional" spin-valve stack, NiFe(8)/Cu(4)/CoFe(4)/IrMn(5), in which the pumped spin current from NiFe is entirely absorbed by CoFe and does not reach IrMn.  As shown in Fig.~\ref{fig:LWpeaks}(e,f), an enhancement of linewidth again coincides with the switching of exchange-biased CoFe~\cite{Note2}.
The fact that the clear peaks in linewidth are still present -- even though the pumped spin current does not enter IrMn -- rules out any contribution from N\'eel-vector-dependent spin decoherence. 
Instead, the plausible mechanism is linewidth broadening of NiFe FMR caused by nonuniform dipolar fields from CoFe when it breaks up into multiple domains during switching, as illustrated in Fig.~\ref{fig:LWpeaks}(a,d). 
Such FMR linewidth broadening induced by inhomogeneous fields is well known ~\cite{Platow1998,McMichael2003,Twisselmann2003}, although it is typically discussed in the context of the zero-frequency linewidth $\DelH_0$ rather than a peak at a finite frequency as observed here.  
Our results in Fig.~\ref{fig:LWpeaks}, along with those in Figs.~\ref{fig:angle} and \ref{fig:Gilbert}, thus reveal that spin decoherence in the AFM IrMn spin sink is remarkably insensitive to the orientation or uniformity of the N\'eel vector.

Since we find spin decoherence in IrMn to be independent of its magnetization state, this polycrystalline AFM can effectively be modeled as a nonmagnetic spin sink, in which spin decoherence due to spin-orbit coupling is isotropic and parameterized by the spin diffusion length \lsd~\cite{Tserkovnyak2002, Manchon2017}. Although the scatter in the experimental data for $\alpha$ (Fig. ~\ref{fig:Gilbert}(d)) makes precise quantification difficult, analyses with simple models~\cite{Boone2013, Ghosh2012} point to \lsd in the range of $\lesssim$1 nm to $\approx$3 nm, as shown in the Supplemental Material. The relatively short \lsd is consistent with strong spin-orbit coupling in IrMn due to the presence of heavy element Ir. The strong spin-orbit coupling is also corroborated by reports of large spin-Hall effect in IrMn ~\cite{Zhang2014,Mendes2014,Tshitoyan2015,Saglam2018}. We also remark that a recent experiment shows the efficiency of charge-to-spin conversion (spin-orbit torque) to be independent of exchange-bias-pinned antiferromagnetic order in NiFe/IrMn bilayers~\cite{Saglam2018}.

It should be noted that our study focuses on spin transport in one specific type of AFM (polycrystalline Ir$_{20}$Mn$_{80}$). 
An open question, which we will address in a future study, is whether anisotropic spin dephasing can be observed in AFMs with weaker spin-orbit coupling. 
Another remaining question is the role of structural disorder, e.g., grain boundaries, on spin decoherence in AFMs. 
Also, since our results are obtained with \emph{electron}-mediated spin transport through the diamagnetic spacer, we cannot rule out the possibility of significant anisotropic spin decoherence when \emph{magnon}-mediated spin transport dominates, as may be the case for directly exchange-coupled bilayers of FM-spin-source/AFM-spin-sink~\cite{Moriyama2017}. 
However, for spin pumping experiments on FM/AFM bilayers, one would need to  disentangle the anisotropic decoherence of magnon-mediated spin current in the AFM spin sink from the anisotropic relaxation of magnetization dynamics within the FM spin source (e.g., due to two-magnon scattering)~\cite{McMichael1998,Mewes2010,BeikMohammadi2017}. 

In summary, we have examined the decoherence of electronic spin current in polycrystalline AFM IrMn with globally pinned antiferromagnetic order. 
We find that spin decoherence is independent of the relative orientation between the pumped spin polarization and the pinned N\'eel vector. Spin decoherence is also identical for samples with and without pinned antiferromagnetic order in IrMn. 
Our findings highlight the need to further investigate the interplay between spin current and antiferromagnetic order.

\vspace{10pt}
We acknowledge helpful discussion with V. Baltz.


\end{document}